\documentclass[preprint2,usenatbib]{aastex6}
\usepackage{natbib,amsmath}

\shorttitle{Magnetars in ULGRBs}
\shortauthors{Gompertz \& Fruchter}

\begin{document}

\title{Magnetars in ultra-long gamma-ray bursts}

\author{B. Gompertz* and A. Fruchter}
\affil{Space Telescope Science Institute, Baltimore, MD 21218, USA}

\email{*bgompertz@stsci.edu}

\begin{abstract}
Supernova 2011kl, associated with the ultra-long gamma-ray burst (ULGRB) 111209A, exhibited a higher-than-normal peak luminosity, placing it in the parameter space between regular supernovae and super-luminous supernovae. Its light curve can only be matched by an abnormally high fraction of $^{56}$Ni that appears inconsistent with the observed spectrum, and as a result it has been suggested that the supernova, and by extension the gamma-ray burst, are powered by the spin-down of a highly magnetised millisecond pulsar, known as a magnetar. We investigate the broadband observations of ULGRB 111209A, and find two independent measures that suggest a high density circumburst environment. However, the light curve of the GRB afterglow shows no evidence of a jet break (the steep decline that would be expected as the jet slows due to the resistance of the external medium) out to three weeks after trigger, implying a wide jet. Combined with the high isotropic energy of the burst, this implies that only a magnetar with a spin period of $\sim$ 1~ms or faster can provide enough energy to power both ULGRB 111209A and Supernova 2011kl.
\end{abstract}
\keywords{}

\section{Introduction}
For most of the past twenty years, gamma-ray bursts (GRBs) have been split into just two classes, long and short, based on a dichotomy observed in their durations and spectral hardness \citep{Kouveliotou93}, with observation strongly pointing towards two physically distinct types of progenitor. Long GRBs (LGRBs) are typically active for anywhere between around two seconds to up to a few hundred seconds, are associated with host galaxies showing active star-formation \citep[cf.][]{Fruchter06,Levesque10c}, and usually (indeed perhaps always) have underlying Type Ic SNe \citep[cf.][]{Hjorth03,Cano13}. In contrast, short GRBs (SGRBs) typically last for less than 2 seconds, are not associated with SNe, and are found in early type as well as late-type hosts \citep{Berger05b,Fong15}. It appears likely that these bursts result from the merger of compact objects (either two neutron stars or a neutron star-black hole pair), and in two cases may have been observed with associated kilonovae \citep{Tanvir13,Yang15}, in which light is emitted by the radioactive decay of neutron-rich material expelled from the merging binary \citep{Li98}. In the last several years, a new class of GRB with emission lasting thousands of seconds has been identified. These durations are far in excess of those observed in traditional LGRBs \citep{Levan14}. Now called ultra-long GRBs (ULGRBs), their durations are apparently statistically distinct \citep{Boer15,Levan15} from LGRBs.

The physical mechanism and progenitors responsible for ULGRBs is so far unclear. However, a recent analysis of observations of ULGRB 111209A by \citet{Greiner15} has provided a striking clue. In the light curve and optical spectra of ULGRB 111209A, there is a strong suggestion of an underlying, very powerful supernova, SN 2011kl. While the Type Ic SNe of normal LGRBs appear to be powered by the decay of $^{56}$Ni \citep[cf.][]{Cano16}, SN 2011kl appears to show a total energy in excess of what could be attributable to $^{56}$Ni unless it comprised a dominant fraction of the ejecta. Instead, \citet{Greiner15} suggest that the supernova is primarily driven by spin-down radiation from a highly magnetized millisecond pulsar, known as a magnetar. The suggestion that GRBs might be associated magnetars is not new. It goes back at least until the early 1990s \citep{Usov92,Thompson95}. However, there has been very little observational evidence which would prefer a GRB powered by a magnetar rather than an accreting black hole. The strongest evidence so far had been a suggestion by \citet{Mazzali14} that an apparent upper limit on the kinetic energy of SNe associated with LGRBs could be explained by magnetars, as the total rotational energy of a neutron star (NS) cannot exceed $\sim 5\times10^{52}$~ergs, since the break-up rotational period is not greatly below 1~ms and the mass not much above $2$~M$_{\sun}$.

This paper investigates the observations of GRB 111209A in an effort to place constraints on the magnetar model. In Section~\ref{sec:energy}, we discuss the various energy requirements of the GRB and the supernova, and compare them to the limits of magnetar spin down. Section~\ref{sec:afterglow} examines the X-ray and optical afterglow observations, and what they can tell us about the GRB and SN environment. Section~\ref{sec:radio} looks at the radio afterglow observations. We discuss the implications for the magnetar model from the implied densities in Section~\ref{sec:discussion} and summarise our conclusions in Section~\ref{sec:conclusions}.

\section{The magnetar energy budget}\label{sec:energy}

The suggestion that ULGRB 111209A/SN 2011kl is powered by a magnetar has sparked great interest in the community, and a number of authors have proposed magnetar physical parameters (spin periods and dipole field strengths) capable of reproducing the observations \citep{Greiner15,Metzger15,Bersten16,Cano16}. However, many of these have focused on powering the SN alone. Magnetic dipole spin-down draws its energy from the rotational reservoir of the magnetar, and so using it to power a GRB places strong limits on the total energy available. The limiting factor of the total energy available is the initial spin period of the NS, and the total energy available is
\begin{equation} \label{eqn:energy}
E \approx 1.5\times10^{52}M_{\sun}R^2_{10}P_0^{-2} \mbox{ erg,}
\end{equation}
where $P_0$ is the initial NS spin period in ms, $M_{\sun}$ is the mass of the NS in solar masses and $R_{10}$ is the NS radius in units of $10$~km. Observations of NSs with masses slightly in excess of $2~M_{\sun}$ \citep{Demorest10,Antoniadis13} suggest at least $3\times10^{52}$~erg is attainable for a millisecond magnetar, and \citet{Metzger15} show that in extreme cases, up to $\sim 10^{53}$~erg is attainable for larger NS radii and sub-ms spin periods. Both the SN and GRB benefit from the energy released during core collapse, but while the total gravitational energy released is $\sim 10^{53}$~erg, the majority of this is carried away by neutrinos, leaving of the order of a few $\times 10^{51}$~erg, depending on the mass of the stellar core and the efficiency of the energy conversion. The SN later gains energy from radioactive decay \citep[cf.][]{Bersten16,Cano16}. The GRB has the magnetar contribution, plus energy released during early accretion, which is expected to again be of the order of a few $\times 10^{51}$~erg, assuming the formation of a $2$~M$_{\sun}$ NS that accretes up to $0.1$~M$_{\sun}$ with an energy conversion efficiency in the region of $10$ per cent.

Observations of SN 2011kl taken by \citet{Greiner15} show its spectrum to be rather featureless due to line blending, supporting a photospheric velocity $v_{\rm ph} \gtrsim 20,000$~km~s$^{-1}$. The mean expansion velocity is expected to be related the the photosphere velocity by $\langle v^2 \rangle = 3/5 v_{\rm ph}^2$ \citep{Wheeler15}. Previous studies on SN 2011kl find an ejecta mass $\sim 3M_{\odot}$ from their modelling \citep[e.g.][]{Greiner15,Bersten16}, so a kinetic energy $E_k = \frac{3}{10}M_{\rm ej}v_{\rm ph}^2 \sim 7\times10^{51}$~erg is required. However, if the SN is aspherical to a degree comparable to SN 1998bw, this value may decrease by as much as a factor of 5 \citep[e.g.][]{Maeda02,Tanaka07}. Integrating the light curve of SN 2011kl results in a total luminosity of $\sim 10^{50}$~erg. The entire energy requirement of the SN is therefore at least $\sim 2\times 10^{51}$~erg, so a magnetar with a spin period faster than $3.5$~ms is required to power the SN alone. However, these are idealised conditions, and a spherical SN 2011kl could have a kinetic energy as high as $1.2\times 10^{52}$~erg (depending on the velocity distribution of the ejecta), placing it at similar energies to the GRB SNe in \citet{Mazzali14}. In this case, even a 2~ms magnetar would be unable to provide the energy required by  the SN.

\emph{Konus-Wind}, which is able to observe GRBs without the interruptions caused by low-earth orbit, reported an isotropic equivalent $\gamma$-ray energy release of $E_{\gamma, \rm iso} = (5.7 \pm 0.7)\times10^{53}$~erg \citep{Golenetskii11} from GRB~111209A, placing it towards the brighter end of the distribution of GRB energies \citep[cf.][]{Nava12}. This energy release is subject to two sources of inefficiency, the first being the proportion of energy that makes it into the jet (energy lost here finds its way into the supernova), and the second being the efficiency of the radiation process in the intra-jet shocks that produce the observed $\gamma$-ray emission. Energy lost here ends up in the GRB afterglow. The latter process can be up to 50 per cent efficient \citep{Beniamini15}, meaning that the total isotropic equivalent energy release summed across all wavelengths is at least double the observed value of $E_{\gamma, \rm iso}$.

The true energy release is related to the isotropic equivalent energy by
\begin{equation}
E = (1-\cos \theta) E_{\rm iso} \mbox{ erg,}
\end{equation}
where $\theta$ is the jet half-opening angle in radians. This correction is also often discussed in terms of a beaming factor, which is just $b_f = (1-\cos \theta)^{-1}$. The $E_{\rm iso} \sim 10^{54}$~erg seen in GRB~11209A, therefore, can only be reconciled with a magnetar central engine if the jet had a very tight beam, and thus a large beaming factor. The most energetic model in the literature for ULGRB 111209A/SN 2011kl is that of \citet{Metzger15}, who use a 2~ms magnetar (total energy $7.5\times 10^{51}$~erg). Because they attempt to model the GRB as well as the supernova, the authors are forced to reproduce the high $E_{\rm iso}$, and hence use a high beaming factor of 800, corresponding to a jet opening angle of $\theta_0 = 0.05$~radians. This is at the narrow end of GRB jet opening angles \citep[see e.g.][]{Frail01,Ryan15}. In contrast, \citet{Ryan15} modelled the X-ray light curve of 111209A in an effort to find the jet opening angle independent of any central engine model. They find a jet opening angle of $0.34^{+0.11}_{-0.13}$ radians, corresponding to a beaming factor of $17.5^{+28.0}_{-7.5}$. The \citet{Ryan15} beaming factor is more than $30$ times smaller than that required by the \citet{Metzger15} model.

\subsection{Jet Breaks}\label{sec:breaks}

At first, the observer of a GRB afterglow receives only photons from the shock front of a strongly collimated jet. However, as the expanding jet is slowed by the circum-burst medium (CBM), the beaming lessens, and thus the emission from the shock front is radiated to a rapidly increasing region of the sky, so that an observer will eventually be able to observe the non-isotropic nature of the outflow for the first time \citep[see e.g.][]{Rhoads99}. This manifests in the light curves as an achromatic drop in flux, known as a jet break. The narrower the jet opening angle, the earlier the break should be observed. The very narrow jet employed by \citet{Metzger15} suggests a jet break may occur fairly early in the light curve evolution.

The time at which the jet break should become apparent depends on the radial profile of the CBM.  If the jet is expanding into a stellar wind with an density profile $\propto r^{-2}$, one finds a jet break time of \citep[cf.][]{Chevalier00}
\begin{equation} \label{eqn:wind}
t_j = 10\times(1+z)\bigg(\frac{\theta_0}{0.2}\bigg)^4\eta^{-1} E_{\gamma,\rm iso,53}A_*^{-1} \mbox{ days,}
\end{equation}
where $E_{\gamma, \rm iso,53}$ is the isotropic equivalent $\gamma$-ray energy in units of $10^{53}$~erg, $\eta$ is the fractional efficiency of the $\gamma$ emission in the shocks, and $A = \dot{M}_w/4\pi V_w = 5\times 10^{11}A_*$~g~cm$^{-1}$ is the wind density. For an ISM-like profile where the density is roughly constant with radius, the equation becomes \citep[cf.][]{Racusin09b}
\begin{equation} \label{eqn:ism}
t_j = 2.85\times10^2\, \theta_0^{8/3}(1+z)\bigg(\frac{E_{\gamma,\rm iso,53}}{\eta n_0}\bigg)^{1/3} \mbox{days,}
\end{equation}
where $n_0$ is the number density (atoms per cm$^{3}$) of the CBM.

\emph{Swift}-XRT tracked the burst until $21.9$ days after trigger, which corresponds to $13.1$ days rest frame at the $z = 0.677$ redshift of GRB~111209A \citep{Greiner15}. Nonetheless, the late-time light curve fits a single power-law decay of $1.43 \pm 0.06$ and thus shows no evidence of the sharp turn-over that would be expected in the case of a jet-break. However, to avoid a break before 22 days in the narrow-jet model proposed by \citet{Metzger15} requires that $A_* \le 3.41\times10^{-2}$~g~cm$^{-1}$ in the case of a wind (Equation~\ref{eqn:wind}) or $n_0 \le 4.63\times10^{-6}$~atoms~cm$^{-3}$ in the case of the ISM (Equation~\ref{eqn:ism}), where in both cases we assume $50$ per cent efficiency in $\gamma$-ray emission, as is done by \citet{Metzger15}. The latter case is not much above the intergalactic density, making the homogenous medium solution extremely unlikely. The required wind density is not particularly low, and a 2~ms magnetar can be reconciled with observations if the GRB occurred in a very sparse environment, since the available energy would then be high enough to keep the ejecta at velocities sufficient for Doppler beaming to mask the jetted nature of the outflow out beyond the XRT observations. However, we argue in Section~\ref{sec:afterglow} that this is not the case, and that the environment of GRB~111209A is in fact fairly dense. It should be noted that wide observer angles with respect to the jet can smear out a jet break \citep{vanEerten12}, making them harder to observe. This could be the case in 111209A, based on the ratio of observer angle to jet opening angle $\frac{\theta_{obs}}{\theta_{jet}} = 0.57^{+0.30}_{-0.38}$ found by \citet{Ryan15}, though the uncertainty is large. Nonetheless, there is no evidence of any curvature in the X-ray temporal decay that might signal an approach to values that could be considered post jet-break ($\alpha_x \gtrsim 2$).

\section{The afterglow of 111209A}\label{sec:afterglow}

\begin{figure}
\begin{center}
\includegraphics[width=8.5cm]{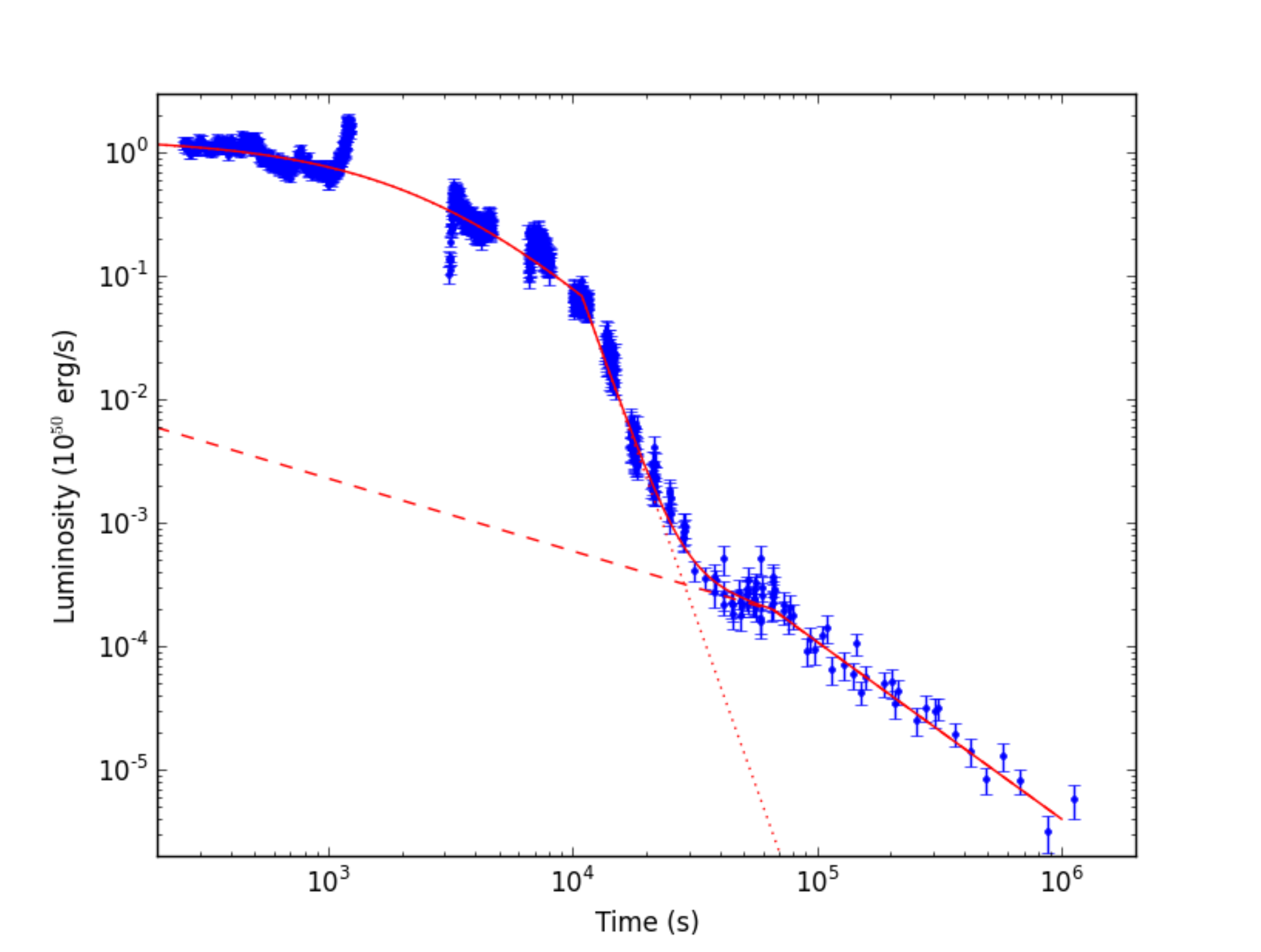}
\end{center}
\caption{\emph{Swift}-XRT observations of ULGRB 111209A corrected to the rest frame. The data are fitted with a dipole spin-down model \citep[cf.][]{Zhang01} to test the model's compatibility, followed by a broken power law to represent the afterglow. The early part of the light curve is driven by the jet, followed by a steep drop close to $10^4$~s due to a loss of collimation \citep[e.g.][]{Metzger11}. The later part represents the forward shock afterglow, completing the two emission component model. No potential jet break is visible out to $21.9$ days in the observer frame. \label{fig:X-ray}}
\end{figure}

\emph{Konus-Wind} observed episodes of $\gamma$ emission from GRB 111209A out to $\sim 10,000$~s after the \emph{Swift}-BAT trigger, at which time a rapid drop in flux was observed in the \emph{Swift}-XRT light curve (Figure~\ref{fig:X-ray}) from the UK \emph{Swift} Science Data Centre \citep[UKSSDC;][]{Evans07,Evans09}. This drop is too steep to be afterglow emission, and strongly indicates that the preceding X-ray light curve was driven by an internal process. An explanation for this behaviour in the protomagnetar model \citep{Metzger11} is that the prompt emission jets are collimated by the stellar envelope, and are driven by the magnetar until the proto-NS becomes transparent to neutrinos, at which point magnetisation rapidly increases and jet acceleration becomes ineffective. As a result, the jet ceases, producing the rapid drop. We fit the early X-ray plateau with the dipole spin-down model of \citet{Zhang01} (see Figure~\ref{fig:X-ray}) to test the model's compatibility with the data, and include a break leading to the rapid decay around $10^4$~s, which we assume to be caused by the decoupling of the jet and dipole due to the loss of collimation \citep[e.g.][]{Metzger11}. The dipole model predicts a flat plateau segment at times earlier than the characteristic spin-down timescale ($T_{\rm em}$) of the magnetar, and a $t^{-2}$ power law at times later than this. The sloping curvature of the early X-ray emission indicates that $t$ is approaching $T_{\rm em}$, where the two are smoothly joined by a gradual turnover in flux. The luminosity of the plateau ($L_{\rm em,0}$) and $T_{\rm em}$ take the form
\begin{equation}
T_{\rm em} \approx 2.05\times10^3 I_{45}B_{p,15}^{-2}P_{0}^2R_{10}^{-6} \mbox{ s}
\end{equation}
and
\begin{equation}
L_{\rm em,0} \approx 10^{49}f_bB_{p,15}^2P_{0}^{-4}R_{10}^6 \mbox{ erg~s$^{-1}$,}
\end{equation}
where $I_{45}$ is the moment of inertia of the NS in $10^{45}$~g~cm$^2$ and $B_{p,15}$ is its magnetic dipole field strength in units of $10^{15}$~G. These parameters match the data for dipole spin-down models with spin periods faster than 3~ms, with the main variation between the fits being the dipole field strength and beaming factor ($f_b$). This ranges from $B \sim 2.5 \times 10^{15}$~G, $f_b \sim 520$ for a 3~ms magnetar to $B \sim 8.3 \times 10^{14}$~G, $f_b \sim 60$ for a 1~ms spin period. The $1$~ms plot is shown in Figure~\ref{fig:X-ray}. These values of $B$ and $P$ assume an NS radius of $10$~km and a moment of inertia of $10^{45}$~g~cm$^2$.

The X-ray light curve at times greater than $6\times10^4$~s (i.e. directly after the rapid decay) is best fit with a broken power law, featuring a break at $(1.12\pm0.2)\times10^5$~s and a final power law decay of $\alpha_x = 1.43\pm0.06$ ($F \propto t^{-\alpha}$). This is compatible within errors with the values found for the same power-law segment by \citet{Gendre13} and \citet{Stratta13}, who found $\alpha_x = 1.51 \pm 0.08$ and $\alpha_x = 1.52 \pm 0.06$, respectively. Our result is steeper than the index found by \citet{Levan14} ($\alpha_x = 1.36 \pm 0.05$), who did not include a break at $\sim 10^5$~s, though our measured $\alpha_x$ is still compatible within errors with their findings. The X-ray spectrum beyond $10^5$~s is best fit with a power law with an index of $\beta_x = 1.50\pm0.12$ ($F \propto \nu^{-\beta}$) and absorption of $2.5^{+0.67}_{-0.61}$~$\times 10^{21}$~cm$^{-2}$ (consistent with the automatic \emph{Swift} spectra) above the galactic value of $1.54 \times 10^{20}$~cm$^{-2}$, using the time-slice spectrum tool from the UKSSDC. All errors here are $1\sigma$. The need for a broken power law in GRB afterglows is often due to the transition from the shallow decay phase, where the afterglow is still being re-energised by the internal emission, to the normal decay phase, which is a pure afterglow decay, as per the `canonical' GRB X-ray afterglow \citep{Nousek06,OBrien06,Zhang06}. However, in 111209A, any shallow decay plateau is hidden beneath the longer-than-normal prompt emission, and the flattening prior to $\sim 10^5$~s appears to be related to some kind of flaring activity, as evidenced by a chromatic brightening in optical and IR bands \citep{Kann16}. After the break, the X-ray afterglow very closely follows a single power-law from around $1$ to $22$ days; there is no sign of curvature, and therefore no evidence of a jet break.

\subsection{Optical Afterglow}\label{sec:optafterglow}

\citet{Greiner15} observed the afterglow of ULGRB 111209A with GROND \citep{Greiner08}, obtaining simultaneous images in the optical g', r', i' and z' bands, and near infra-red (nIR) J, H, and K bands over 20 epochs. The first 9 of these are sufficiently early so as to not be affected by the rising supernova, and also have roughly contemporaneous X-ray observations. The temporal slopes in each optical/nIR band are highly consistent, with a mean of $\alpha_o = 1.30 \pm 0.05$, in agreement with \citep{Levan14}. Each SED was given an arbitrary normalisation, which allowed us to fit all 9 together to obtain better fit statistics. After correcting for the small amount of Galactic absorption \citep[$E(B-V) = 0.01$][]{Schlafly11}, the combined SEDs fit a simple power law spectral slope $\beta_o = 1.23\pm0.02$. This index would result in the optical fluxes falling well below the X-ray flux measured by \emph{Swift}-XRT when extrapolated to the XRT bandpass. The synchrotron spectrum must be connected, and we therefore fitted the 9 SEDs simultaneously to a power-law multiplied by the parameterised extinction curves of \citet{Cardelli89} to include the effect of intrinsic absorption by the host galaxy. Our best fit was $\beta_o = 1.08\pm0.07$ with a rest frame V-band extinction magnitude of $A_V = 0.10 \pm 0.05$. This measured $\beta_o$ is almost identical to the value of $1.07 \pm 0.15$ found for the single SED at $T_0 + 63$~ks by \citet{Stratta13}, though their result is obtained with no intrinsic extinction. Our result is compatible with each of their findings for $\beta_o$ during the optical afterglow. The best fitting extinction law model was the SMC, with $R_v = 2.74$ \citep{Gordon03}, but the fits for the Milky Way ($R_v = 3.1$) and LMC ($R_v = 3.41$) provided almost as good fit statistics with highly similar results.

\subsection{Density Implications from the SED}

\begin{figure}
\begin{center}
\includegraphics[width=8.5cm]{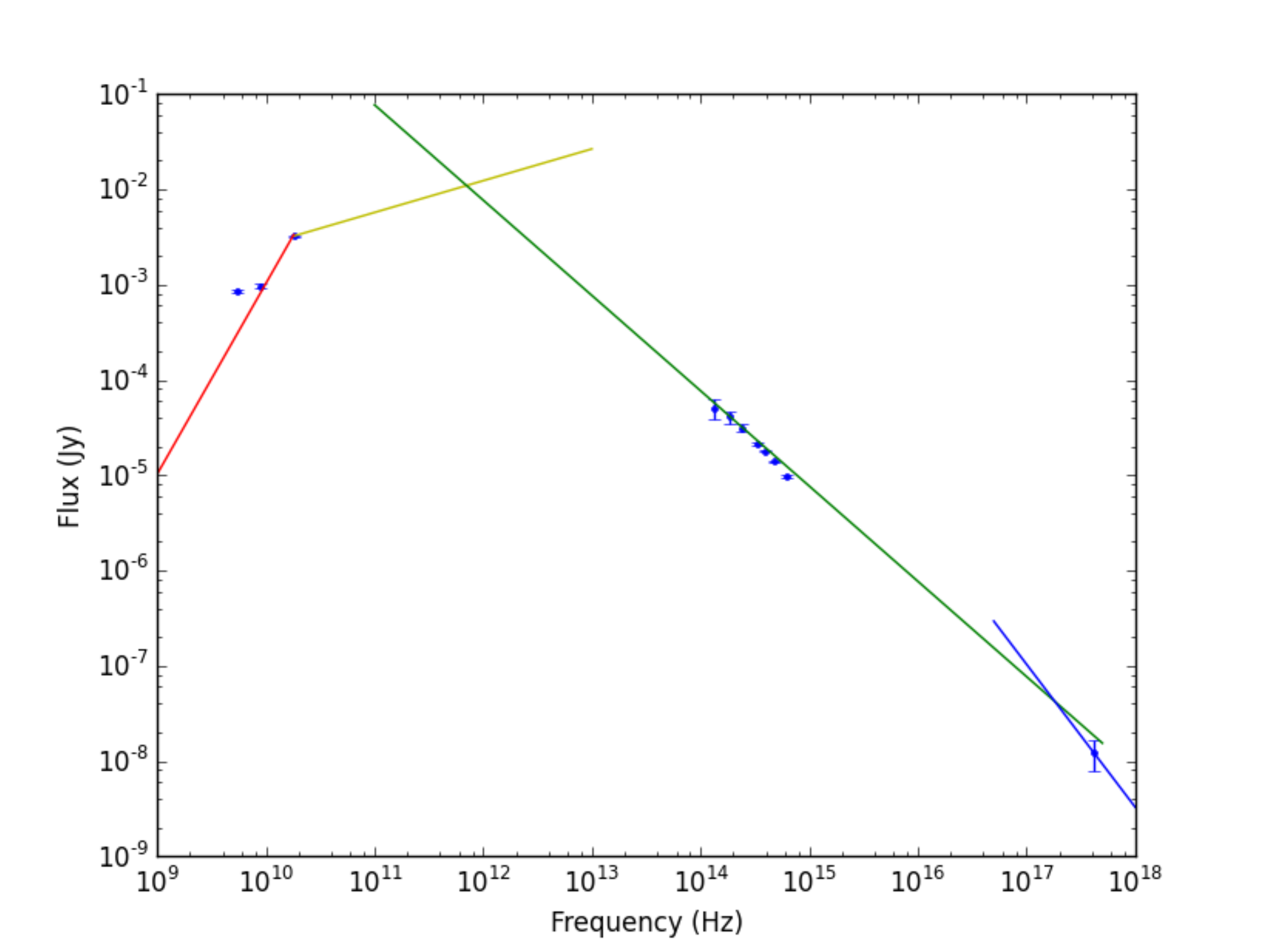}
\end{center}
\caption{A toy model SED showing the synchrotron spectrum for the available observations close to 5 days after trigger. The disjoint between the nIR/optical index (green) and the X-ray index (blue) shows that the synchrotron cooling break must lie below the X-ray frequency. The radio observations (the three points around $10^{10}$~Hz), their compatibility with the synchrotron spectrum, and the position of the synchrotron self-absorption break (connecting the red and yellow segments) are discussed in Section~\ref{sec:radio}. \label{fig:SED}}
\end{figure}

Despite being measured independently of one another, the X-ray and optical spectral indices of $\beta_x = 1.50 \pm 0.12$ and $\beta_o = 1.08 \pm 0.07$ differ by very close to $0.5$ ($\beta_x - \beta_o = 0.42 \pm 0.14$), which is the offset predicted for a synchrotron cooling break, where the synchrotron cooling frequency, $\nu_c$, lies between the two bands \citep[see e.g.][]{Gao13}. Similarly, the temporal indices of $\alpha_x = 1.43 \pm 0.06$ and $\alpha_o = 1.30 \pm 0.05$ also strongly suggest a cooling break, though they are a little further away from the theoretically expected difference of $0.25$ ($\alpha_x - \alpha_o = 0.13 \pm 0.08$). The predicted offset is even more closely recovered if the X-ray temporal indices of \citet{Gendre13} ($\alpha_x = 1.51 \pm 0.08$) or \citet{Stratta13} ($\alpha_x = 1.52 \pm 0.06$) are adopted. The need for a cooling break between the bands is highlighted in Figure~\ref{fig:SED}, where the well-measured X-ray spectral index would greatly over-estimate the observed optical flux if no break is present. Placing $\nu_c$ below the XRT bandpass lower limit $\nu_x$ helps to constrain the physical parameters of the environment.
\begin{equation}
\nu_c = 3.15\times10^{12}\epsilon_{B,-1}^{-3/2}A_*^{-2}E_{\rm k,iso,53}^{1/2}t_d^{1/2}\bigg(\frac{1+z}{2}\bigg)^{-3/2} \mbox{Hz}
\end{equation}
for the wind case, and
\begin{equation}
\nu_c = 1.89\times10^{13}\epsilon_{B,-1}^{-3/2}n_0^{-1}E_{\rm k,iso,53}^{-1/2}t_d^{-1/2}\bigg(\frac{1+z}{2}\bigg)^{-1/2} \mbox{Hz}
\end{equation}
for the ISM \citep[cf.][]{vanderHorst07}. $\epsilon_{B,-1} = \epsilon_B / 0.1$, where $\epsilon_B$ is the fraction of the available energy that is contained in the magnetic fields. $E_{\rm k, iso, 53}$ is the total kinetic energy (in units of $10^{53}$~erg), assuming an isotropic explosion, or nearly equivalently, the total energy radiated by the afterglow (under the assumption of isotropy). For both environment types, $\nu_c \leq 7.25\times10^{16}$~Hz, which is the X-ray frequency at $0.3$~keV; the lower limit of the \emph{Swift}-XRT bandpass. We therefore obtain the limits
\begin{equation}
\begin{aligned}
7.25\times10^{16} \geq 3.15\times10^{12}\epsilon_{B,-1}^{-3/2}A_*^{-2}E_{\rm k,iso,53}^{1/2}t_d^{1/2}\\ \times \bigg(\frac{1+z}{2}\bigg)^{-3/2} \mbox{Hz}
\end{aligned}
\end{equation}
for the wind case, and
\begin{equation}
\begin{aligned}
7.25\times10^{16} \geq 1.89\times10^{13}\epsilon_{B,-1}^{-3/2}n_0^{-1}E_{\rm k,iso,53}^{-1/2}t_d^{-1/2}\\ \times \bigg(\frac{1+z}{2}\bigg)^{-1/2} \mbox{Hz}
\end{aligned}
\end{equation}
for the ISM, which can be rearranged to place limits on the density:
\begin{equation}
\begin{aligned}
A_* \geq 6.59\times10^{-3}\epsilon_{B,-1}^{-3/4}\big((1-\eta)E_{\rm iso,53}\big)^{1/4}t_d^{1/4}\\ \times \bigg(\frac{1+z}{2}\bigg)^{-3/4} \mbox{ g~cm$^{-1}$,}
\end{aligned}
\end{equation}
and
\begin{equation}
\begin{aligned}
n_0 \geq 2.61\times10^{-4}\epsilon_{B,-1}^{-3/2}\big((1-\eta)E_{\rm iso,53}\big)^{-1/2}t_d^{-1/2}\\ \times \bigg(\frac{1+z}{2}\bigg)^{-1/2} \mbox{ atoms cm$^{-3}$.}
\end{aligned}
\end{equation}
Note that in these last two equations, we have substituted $E_{\rm k,iso,53} = (1-\eta)E_{\rm iso,53}$ to highlight the role of efficiency; a lower efficiency in the prompt emission ($\eta$) means that more of the total energy ($E_{\rm iso}$) is available as energy in the afterglow ($E_{\rm k,iso}$). Assuming a $50$ per cent efficiency, $E_{\rm k,iso} = E_{\rm \gamma,iso} = 5.7\times10^{53}$, and the redshift of GRB 111209A is $z = 0.677$. $\nu_c$ is less than $\nu_x$ throughout the evolution of the afterglow in Figure~\ref{fig:X-ray}, so we can choose the most constraining times (21.9~d in the wind case and 1~d for the ISM) to find useful limits on the density as a function of $\epsilon_B$. These limits are
\begin{equation}
A_* \geq 2.50\times10^{-2}\epsilon_{B,-1}^{-3/4} \mbox{ g~cm$^{-1}$}
\end{equation}
and
\begin{equation}
n_0 \geq 1.19\times10^{-4}\epsilon_{B,-1}^{-3/2} \mbox{ atoms cm$^{-3}$.}
\end{equation}
Since 50 per cent is likely the upper limit of the efficiency \citep[e.g.][]{Beniamini15}, it therefore represents the true lower limit of the density in the wind case. For the ISM environment, the minimum density required to keep $\nu_c$ below $\nu_x$ goes down as energy increases, and for a prompt emission jet that was, for example, only $1$ per cent efficient, the density lower limit drops to $n_0 \geq 1.20\times 10^{-5}\epsilon_{B,-1}^{-3/2}$~atoms~cm$^{-3}$. This corresponds to an isotropic equivalent energy of $5.64\times10^{55}$~erg. However, even with this concession, this lower density limit is inconsistent with the maximum density permitted to avoid a jet-break during the course of the X-ray light curve for a tight beam with a beaming factor of $800$, as we calculated for the \citet{Metzger15} model ($n_0 \leq 4.63 \times 10^{-6}$~atoms~cm$^{-3}$, Section~\ref{sec:breaks}). The wind case can still be consistent  with our limit of $A_* \leq 3.41\times10^{-2}$~g~cm$^{-1}$ from Section~\ref{sec:breaks} if $\epsilon_B \geq 0.07$. $\epsilon_B$ is often found to be significantly below $0.1$ \citep[e.g.][]{BarniolDuran14,Beniamini15}, which provides a higher density limit, but has also been found at values close to equipartition \citep{Cenko10,Cenko11}. The equipartition value of $\epsilon_B$ is $0.33$; almost $5$ times higher than the lower limit, and so we cannot rule out the narrow beam from \citet{Metzger15} for a wind medium based on the density limits from $\nu_c$ alone.

\section{Clues from the Radio Afterglow}\label{sec:radio}

The radio afterglow of ULGRB 111209A was detected by the Australia Telescope Compact Array (ATCA) at $5.1$~d after the GRB trigger \citep{Hancock12}. The measured flux densities were $0.85\pm0.04$~mJy at $5.5$~GHz, $0.97\pm0.06$~mJy at $9$~GHz, and $3.23\pm0.05$~mJy at $18$~GHz. This emission must be transient in nature due to the non-detection at $1.9$~d down to $132$~$\mu$Jy at $34$~GHz \citep{Hancock11}; in the radio and mm, which lie below the peak frequency of the synchrotron spectrum (Figure~\ref{fig:SED}), the flux varies with frequency as $F \propto \nu^{1/3}$, which means that a $132$~$\mu$Jy upper limit at $34$~GHz corresponds to a $107$~$\mu$Jy upper limit at $18$~GHz. This is over an order of magnitude below the flux observed $3.2$~days later. At very low frequencies, this behaviour changes to $F \propto \nu^2$ when the observed frequency lies below the synchrotron self-absorption frequency, $\nu_a$. This is the threshold at which photons are strongly scattered by the electrons in the medium from which they are emitted, so that they cannot escape to the observer and are simply exchanged internally. All of the observed emission in this spectral regime comes from a thin layer at the surface of the emission site, and the flux is reduced accordingly.

\begin{figure}
\begin{center}
\includegraphics[width=8.5cm]{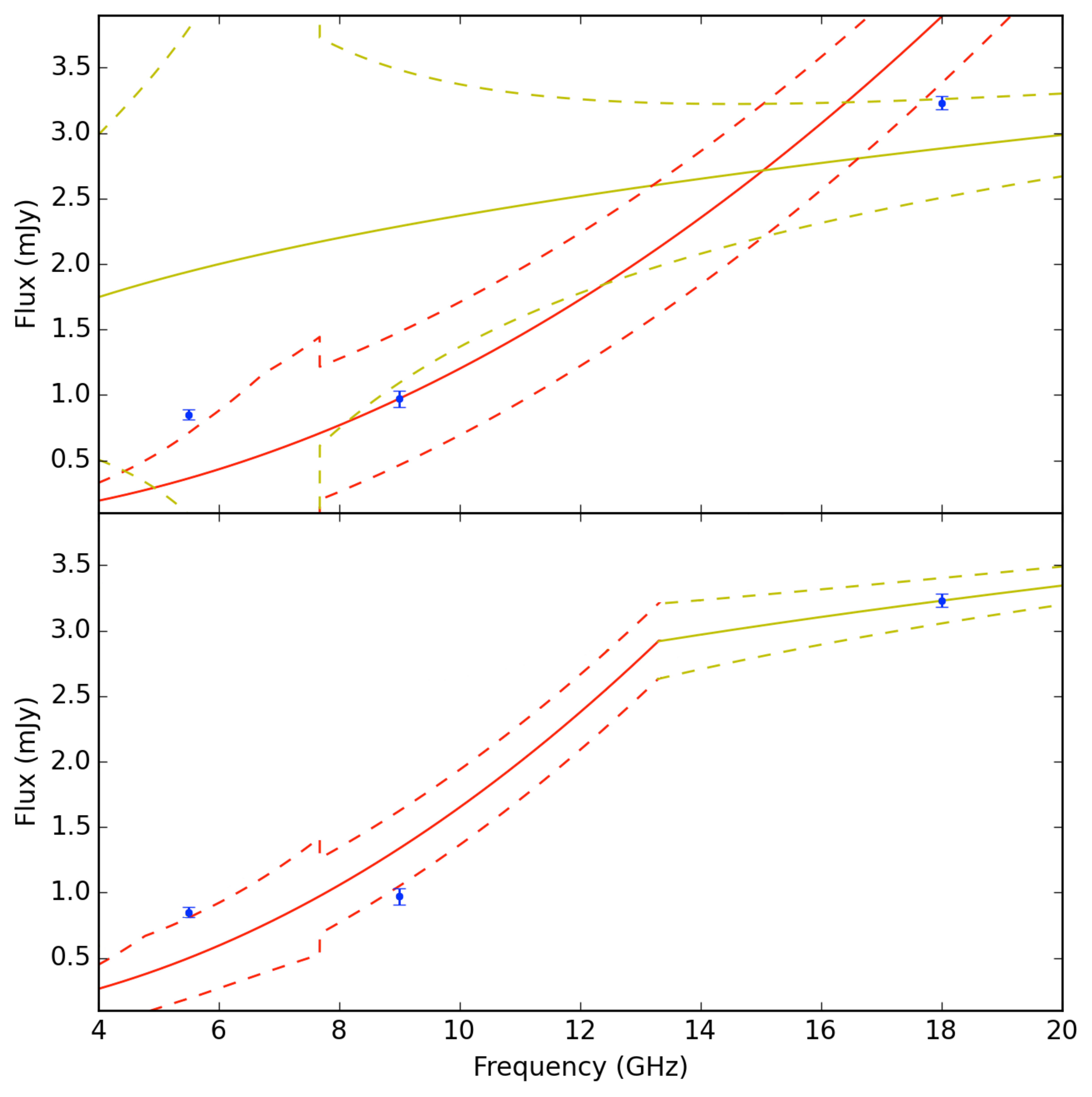}
\end{center}
\caption{Radio detections of the afterglow of ULGRB 111209A at $5.1$~d after trigger. The solid red line shows the model for $\nu$ less than $\nu_a$, the synchrotron self-absorption frequency. The solid yellow line is for $\nu$ greater than $\nu_a$. The dashed lines indicate the amplitude of variability achievable at each frequency due to interstellar scintillation. The top panel shows the amplitude of scintillation required when there is no spectral break; the data must be explained by \emph{either} the red lines \emph{or} the yellow lines. This demands a compact source so that a large amplitude of scintillation can be produced; the source size used for the plot is $4$~$\mu$as ($8.80 \times 10^{16}$~cm). A high density is therefore required to limit the growth of the fireball. The lower panel allows a spectral break, so that the lower frequencies can be matched by the red line, and the higher frequencies by the yellow. The two meet at $\nu_a$, which therefore must be above $9$~GHz. This demands a smaller amplitude of scintillation, which lessens the required compactness. A source that is $10$~$\mu$as ($2.20\times10^{17}$~cm) is sufficient, and the CBM density can therefore be less. \label{fig:scin}}
\end{figure}

The radio detections at three different frequencies in a single epoch from \citet{Hancock12} allow us to investigate which spectral regime presides; $F \propto \nu^{1/3}$ (``unabsorbed") or $F \propto \nu^{2}$ (``absorbed"). We find that between $5.5$ and $9$~GHz, $F \propto \nu^{1/3}$ provides a better match to the observed fluxes, but that between $9$ and $18$~GHz, the observations are best described by $F \propto \nu^2$ (see Figure~\ref{fig:scin}). This is the opposite of what is expected, because the steeper index ought to be at lower frequencies for a synchrotron spectrum with a self-absorption break. Both models provide very poor fits to the data, though the absorbed model fit is better; $\chi^2 = 191$ in the absorbed model, vs. $840$ for the unabsorbed model, with two degrees of freedom. From this, it's clear not only that a single power-law fit isn't the whole picture for the radio afterglow of GRB 111209A, but that there is more to the observations than the simple multiple power law synchrotron spectrum.

\subsection{Interstellar Scintillation}

One possible explanation for the observed deviation from the synchrotron spectrum predicted by theory is interstellar scintillation. At radio frequencies, photons are subject to refraction and diffraction by free electrons in the Milky Way Galaxy, which can lead to fluctuations in the observed brightness of a given source. Scintillation has been observed in several GRBs, one of the best studied being GRB 970508 \citep{Frail97,Waxman98}, where a large flux variability was measured at early times in the radio band, and was subsequently shown to reduce in amplitude with time. This is exactly what is predicted for a source that is initially compact but rapidly expands, as in the fireball model of \citet{Blandford76}, since the increase in the number of sight lines as the source becomes resolved smoothes out the random variability of the electron screen. Interstellar scintillation is therefore a very natural explanation for any observed variability in GRBs at early times and at radio frequencies.

There are two main types of scintillation: weak scattering, where the focal length of the dominant scatterers in the electron screen is greater than the distance from the observer to the screen, so that the image can be displaced but not multiplied/divided, and strong scattering, where the distance to the screen is greater than the focal distance, so multiple images can be produced. See \citep{Goodman97} for a more detailed explanation of the two. Strong scattering can be further subdivided into diffractive scattering, which is due to the superposition of various scattered light travel paths that cannot be resolved by the telescope, and refractive scattering, which is the random magnification of an individual image on one of these paths. Strong scattering occurs when the characteristic diffraction scattering angle $\theta_d$ is greater than the Fresnel angle, given by \citep[cf.][]{Goodman97}
\begin{equation} \label{eqn:fres}
\theta_F = 8.13\nu^{-1/2}d_{\rm scr,kpc}^{-1/2} \mbox{ $\mu$as,}
\end{equation}
where $\nu$ is the observing frequency, and $d_{\rm scr,kpc}$ is the distance to the electron scattering screen in kpc. In terms of frequency, strong scattering occurs below the transition frequency $\nu_0$, given by \citep[cf.][]{Goodman97}
\begin{equation} \label{eqn:trans}
\nu_0 = 10.4(SM_{-3.5})^{6/17}d_{\rm scr,kpc}^{5/17} \mbox{ GHz,}
\end{equation}
where $SM_{-3.5} = SM / 10^{-3.5}$~m$^{-20/3}$~kpc is the scintillation measure. Equation~\ref{eqn:fres} and Equation~\ref{eqn:trans} can be combined to eliminate $d_{\rm scr,kpc}$ and express the Fresnel angle in terms of $\nu$, $\nu_0$ and SM:
\begin{equation} \label{eqn:newfres}
\theta_F = 4.36\times10^2 (SM_{-3.5})^{3/5} \nu^{-1/2} \nu_0^{-17/10} \mbox{ $\mu$as.}
\end{equation}
Both $\nu_0$ and SM can be obtained from the free electron distribution model of \citet{Cordes02}\footnote{online calculator available at http://www.nrl.navy.mil/rsd/RORF/ne2001/}. Using the position on the sky of ULGRB 111209A (Galactic (l,b), (299.92, -70.29)), we obtain $SM = 10^{-3.75}$~m$^{-20/3}$kpc, and a transition frequency between strong and weak scintillation of $7.68$~GHz. To be thorough, we use an integration distance of $100$~kpc to account for electrons across the entire galaxy, though the results are insensitive to this choice above about $4$~kpc due to the large inclination of the source from the Galactic plane. Setting $\nu = \nu_0$ in Equation~\ref{eqn:newfres} gives the Fresnel angle associated with $\nu_0$. This allows us to calculate the amplitude of the three types of scintillation using the equations in Table 1 from \citet{Granot14}. Diffractive scintillation is a narrow-band phenomenon, and there is a suppression factor of $(\Delta \nu_{obs} / \nu)^{-1/2}(\nu/\nu_0)^{17/10}$ \citep{Granot14} for cases where the observing bandwidth $\Delta \nu_{obs}$ is greater than the de-correlation bandwidth $\Delta \nu_{dc} = \nu(\nu/\nu_0)^{17/5}$ because the brightness variations are averaged over a large frequency range and cannot be individually resolved. The bandwidth of the ATCA observations is $2$~GHz for all observing frequencies, and the decorrelation bandwidth equals this value at around $5.6$~GHz. Diffractive scintillation therefore begins to be suppressed below this frequency.

Each type of scintillation has a characteristic time scale, which is the time it takes for the random variability to be averaged out. Scintillation will only be observable if the integration time of the observations is much less than this. The ATCA archive\footnote{http://atoa.atnf.csiro.au/} shows the integration times to be $20$ minutes for the observations at $5.5$ and $9$~GHz, and $14.8$ minutes for those at $18$~GHz. Using the equations for the characteristic time scales for refractive, diffractive and weak scattering in Table 1 of \citet{Granot14}, we find $\tau_{\rm ref} = 4.17$~h, $\tau_{\rm diff} = 1.34$~h at $5$~GHz, $\tau_{\rm weak} = 8.31$~h at $9$~GHz, and $\tau_{\rm weak} = 11.8$~h at $18$~GHz; always in excess of the observation times. The characteristic angular scales are also tabulated in Table 1 of \citet{Granot14}. These mark the thresholds beyond which scintillation starts to diminish as the emitting region grows, the effects of which are included in their amplitude equations.

We find that trying to accommodate the radio observations from \citet{Hancock12} in a single spectral regime (absorbed vs. unabsorbed) requires a degree of scintillation that corresponds to an angular source size of at most $\theta_s = 4$~$\mu$as (Figure~\ref{fig:scin}, upper). If $\nu_a$ lies between two of the observations (most likely the ones at $9$ and $18$~GHz), then less scintillation is necessary, and the angular size limit becomes $\theta_s = 10$~$\mu$as (Figure~\ref{fig:scin}, lower).

\subsection{Density Limits from Radio Observations} \label{sec:radiodense}

The magnitude of the observed scintillation is a function of the angular size the source subtends on the sky; as mentioned previously, a small point of light is more susceptible to flux variations as a result of the focusing and defocusing it experiences as it is lensed by free electrons in our Galaxy. As the fireball grows and the source appears larger, the effects of scintillation diminish because the broader sight lines average out the flux variations. The main barrier to the growth of the fireball is the density of the medium into which is expands. We can therefore use the observation of scintillation to limit the source size. This, in turn, places lower bounds on the density of the CBM.

The jet in a GRB is pointed towards the observer, and so the physical source size is the diameter of the opening of the jet. This can be estimated as $d \approx R/\Gamma$~cm, because at early times we assume our view of the opening is of the part of the shock that is beamed towards us, so that $\theta_0 \approx 1/\Gamma$ \citep[cf.][]{Rhoads99}. The radial extent of the jet outwards from the explosion, R, and its bulk Lorentz factor $\Gamma$ can be estimated as \citep[cf.][]{vanderHorst07}:
\begin{equation}
R = (\beta_{\rm ad}E_{\rm iso}t)^{\frac{1}{4-k}}(\alpha_{\rm ad}\rho_0c)^{-\frac{1}{4-k}} \mbox{ cm},
\end{equation}
and
\begin{equation}
\Gamma = (\alpha_{\rm ad}\beta_{\rm ad}^{3-k}\pi c^{5-k}\rho_0E_{\rm iso}^{-1}t^{3-k})^{-\frac{1}{2(4-k)}} \mbox{.}
\end{equation}
These equations are appropriate for an adiabatic expansion. $\rho_0$ is the CBM density (or maximum CBM density for the wind case) in g~cm$^{-3}$. $\alpha_{\rm ad} = 16/(17-4k)$ and $\beta_{\rm ad} = 4-k$ are the adiabatic constants, and k is the density-with-radius index; $k=0$ for the ISM, and $k=2$ for the wind. The source size $d$ can be converted to the angle subtended on the sky using $\theta \approx 206265 \times 10^6 d/D_A$~$\mu$as, where $D_A = 4.62\times10^{27}$~cm is the angular size distance found using a standard cosmology of $H_0 = 67.8$~km~s$^{-1}$~Mpc$^{-1}$, $\Omega_{\rm M} = 0.308$ and $\Omega_{\rm vac} = 0.692$ \citep{PlanckCollaboration16}. The calculation was done with \citet{Wright06}. We now have everything we need to relate the amplitude of scintillation to the density of the environment into which the GRB jet was expanding.

The value of $\theta_s = 4$~$\mu$as, corresponding to the amplitude of scintillation required to explain the radio observations in a single spectral regime, equates to a maximum source size of $8.80\times10^{16}$~cm, which provides density limits of $A_* \geq 70$~g~cm${-1}$ in the wind case, or $n_0 \geq 1.47\times10^4$~atoms~cm$^{-3}$ for an ISM-like environment. If a spectral break is allowed, and $\theta_s = 10$~$\mu$as, the maximum permissible source size is then $2.20\times10^{17}$~cm, and the lower density limit is reduced to $A_* \gtrsim 1.8$~g~cm$^{-1}$ in the wind case, or $n_0 \gtrsim 9.6$~atoms~cm$^{-3}$ for the ISM. However, this second scenario implicitly requires that the synchrotron self-absorption frequency $\nu_a > 9$~GHz, and this carries its own set of density requirements.

In cases like the one at hand, in which the synchrotron breaks are ordered $\nu_a < \nu_m < \nu_c$ (see Figure~\ref{fig:SED}), the synchrotron self-absorption frequency is given by \citep{vanderHorst07}
\begin{equation}
\nu_a = 2.05 \times 10^{11} \epsilon_{e,-1}^{-1} \epsilon_{B,-1}^{1/5} A_*^{6/5} E_{\rm k,iso,53}^{-2/5} t_d^{-3/5} \bigg(\frac{1+z}{2}\bigg)^{-2/5} \mbox{Hz}
\end{equation}
for a wind-like environment, or
\begin{equation}
\nu_a = 1.23 \times 10^{11} \epsilon_{e,-1}^{-1} \epsilon_{B,-1}^{1/5} n_0^{3/5} E_{\rm k,iso,53}^{1/5} \bigg(\frac{1+z}{2}\bigg)^{-1} \mbox{Hz}
\end{equation}
for the ISM, where $\epsilon_{e,-1} = \epsilon_e / 0.1$ is the fraction of the energy contained in the emitting electrons. Requiring that $\nu_a > 9$~GHz means that
\begin{equation} \label{eqn:windlim}
A_* \geq 2.81\times10^{-1} \epsilon_{B,-1}^{-1/6} \mbox{ g~cm$^{-1}$}
\end{equation}
or
\begin{equation} \label{eqn:ismlim}
n_0 \geq 5.33\times10^{-3}\epsilon_{B,-1}^{-1/3} \mbox{ atoms cm$^{-3}$}
\end{equation} 
for $E_{\rm k,iso,53} = 5.7$ and assuming $\epsilon_{e,-1} = 1$. We then have two possible scenarios:\\

1. The radio emission comes from a single power law component (either $F \propto \nu^{1/3}$ or $F \propto \nu^{2}$), and the observed discrepancy from the theoretically expected synchrotron spectrum is due entirely to interstellar scintillation, as in the upper panel of Figure~\ref{fig:scin}. In this case, the density is limited by the maximum source size allowed to produce the amplitude of scintillation required to match the observations.\\

2. The synchrotron self-absorption break $\nu_a$ lies between the observations at $9$ and $18$~GHz. This means that the observations can be matched with a lower amplitude of scintillation, which allows a less compact source size and hence a lower density, but instead places similarly strict limits on the density (shown in Equations \ref{eqn:windlim} and \ref{eqn:ismlim}) because $\nu_a$ must be greater than $9$~GHz.\\

\begin{figure}
\begin{center}
\includegraphics[width=8.5cm]{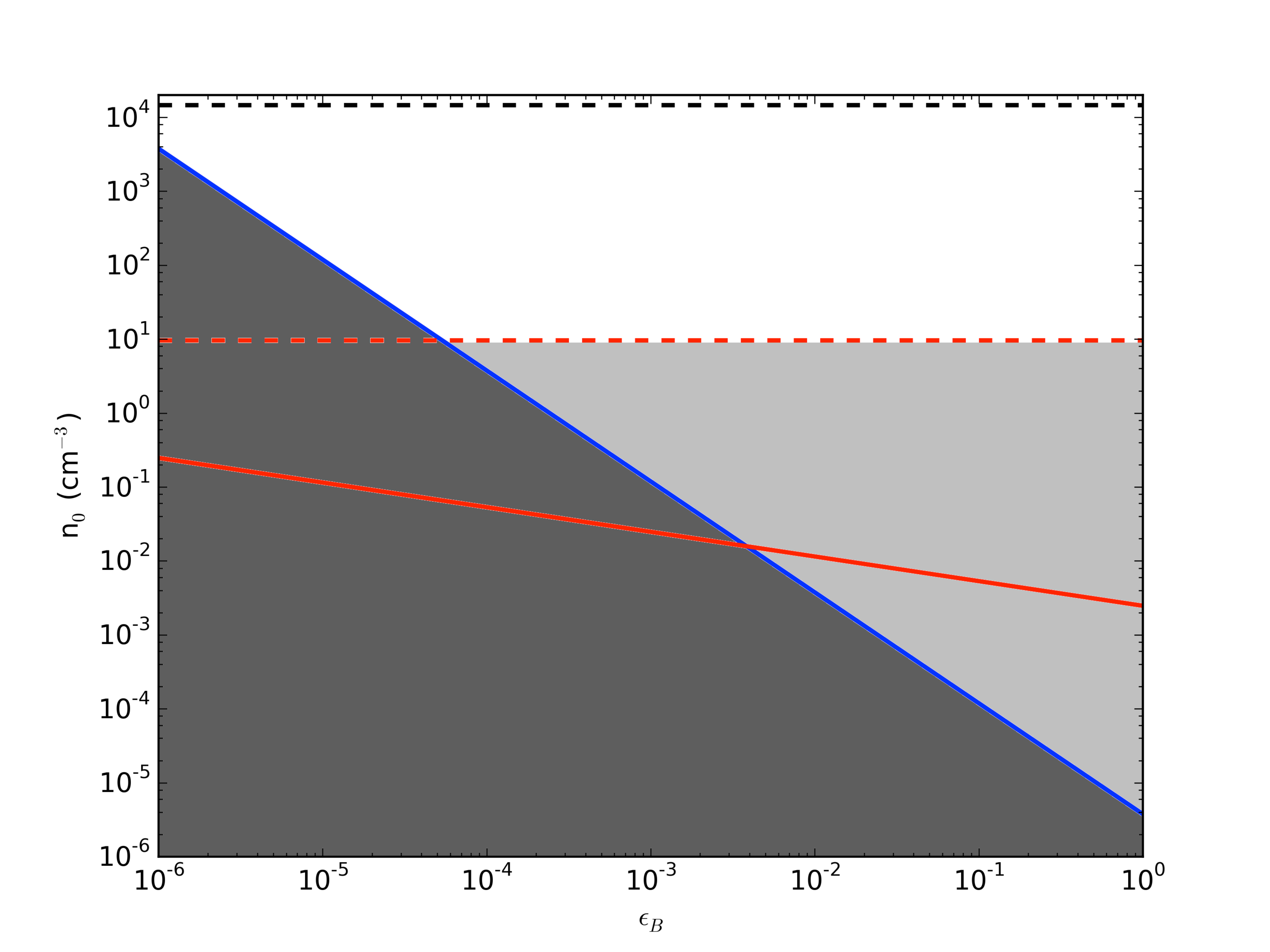}
\includegraphics[width=8.5cm]{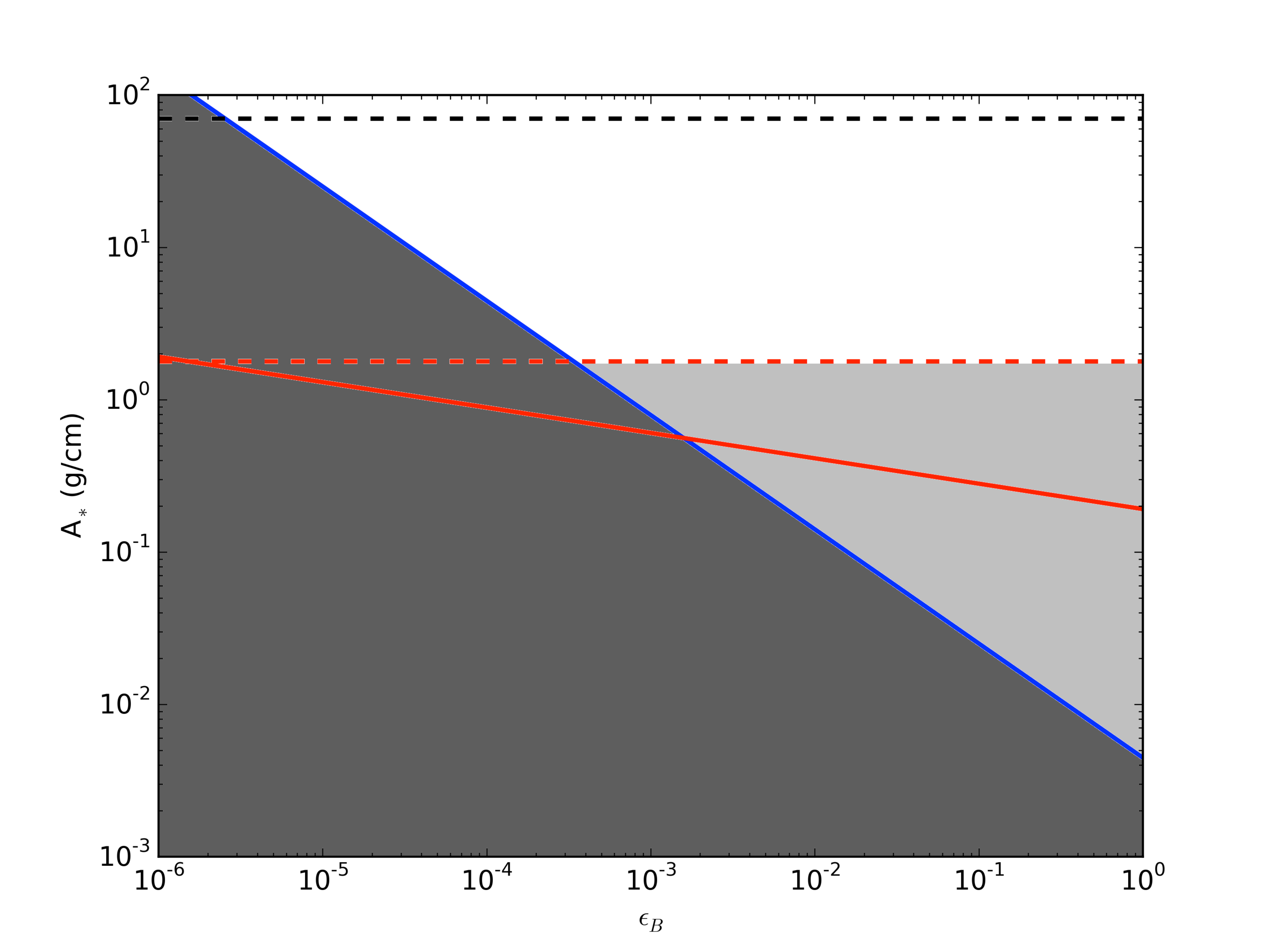}
\end{center}
\caption{Density vs the fraction of afterglow energy contained in the magnetic fields ($\epsilon_B$) for an ISM environment (upper) and wind environment (lower). The blue line marks the values for which the cooling break $\nu_c = 0.3$~keV (the lower limit of the \emph{Swift}-XRT bandpass). The dark grey region is excluded because it requires $\nu_c > 0.3$~keV, whereas the disconnect between the nIR/optical and X-ray spectral indices tells us that $\nu_c $ is between the optical and the X-ray. The black horizontal dashed line marks the density required to restrict the growth of the fireball so that the source is compact enough to provide an amplitude of scintillation sufficient to explain the radio observations in a single spectral regime, as in the top panel of Figure~\ref{fig:scin}. The red horizontal dashed line is the density needed to keep the source compact enough to provide the amplitude of scintillation required if a spectral break is allowed, as in the lower panel of Figure~\ref{fig:scin}. Requiring the spectral break means that $\nu_a \geq 9$~GHz, and this limit is represented by the solid red line. The density must therefore be above both red lines, shown by the white region. If no break is allowed, the density must be above the black line. \label{fig:dense}}
\end{figure}

In the first scenario, the implied wind density of $A_* \geq 70$~g~cm$^{-1}$ or ISM density of $n_0 \geq 1.47 \times 10^4$~atoms~cm$^{-3}$ both seem implausibly high, and likely tell us that a spectral break is required in this part of the spectrum. The wind density is more than a factor of $100$ times higher than anything found in the most energetic GRBs \citep{Cenko10,Cenko11}, and the ISM density is 5 orders of magnitude higher than the densest results of the same study. The second scenario provides more plausible (but still high) results, with the wind density at $A_* \geq 1.79$~g~cm$^{-1}$ and the ISM at $n_0 \geq 9.62$~atoms~cm$^{-3}$. Wind density values this large are associated with Wolf-Rayet stars. The various density limits are illustrated in Figure~\ref{fig:dense}.

\section{Discussion}\label{sec:discussion}

The position of the cooling break and the presence of radio scintillation give us two independent measures that both suggest the CBM density of ULGRB 111209A is high. Figure~\ref{fig:dense} shows the density implications for an ISM or wind-like medium. In both cases, the densities we find are too high to avoid a jet break within the observed X-ray time out to $21.9$~days for a narrow beam with an opening angle of $0.05$~radians, as invoked in \citet{Metzger15}. Taking the minimum densities found in Section~\ref{sec:radiodense} and applying them to Equation~\ref{eqn:wind} and Equation~\ref{eqn:ism} gives us the minimum opening angle required to avoid a jet break: $\theta_0 \geq 0.13$ or $0.31$ for the wind and ISM cases, respectively. These translate into beaming factor upper limits of $b_f \leq 119$ and $21.0$. The beaming factor of $21.0$ associated with the ISM case can only reduce the required energy to $5.4\times10^{52}$~erg, which already tests the available energy from a 1~ms magnetar without even taking the energy required to power SN 2011kl (up to $1.2 \times 10^{52}$~erg; see Section~\ref{sec:energy}) into account, and already assumes an efficiency of $50$ per cent. The beaming factor upper limit in the wind case can reduce the energy demands for the GRB to $E \approx 1\times10^{52}$~erg. This leaves enough energy for even the worst case SN geometry, as described in Section~\ref{sec:energy}, but stretches the energy available from a $2$~ms magnetar, which would then require a $2$~M$_{\sun}$ mass and $12$~km radius NS to be consistent with the GRB alone. However, this pressure may be eased somewhat with a significant contribution from core collapse and early accretion; something at the level of $\gtrsim 3 \times 10^{51}$~erg would provide enough energy when combined with the $7.5 \times10^{51}$~erg budget of a $2$~ms NS. Conversely, if the observations had not ceased, and had continued to show no jet break, the energy demands rise quite rapidly. For the wind environment, $E_{\rm \gamma,iso} \propto t_j$, meaning that another week of unbroken emission would demand $33$ per cent more energy, pushing anything but a magnetar with a spin period close to $1$~ms or below up against its theoretical limits. Indeed, the lack of curvature in the X-ray light curve suggests that no such break is imminent, and the energies discussed here should be considered lower limits. Furthermore, this limit is even more stringent in the ISM case, where $E_{\rm \gamma,iso} \propto t_j^3$.

It is important at this point to recap the assumptions that provide our energy constraints on GRB 111209A. The X-ray light curve in Figure~\ref{fig:X-ray} shows no jet break out to $\sim 22$ days, and we use this fact in Equations~\ref{eqn:wind} and \ref{eqn:ism} to place limits on the minimum opening angle of the jet that would be required to match this observation. We assume a $50$ per cent efficiency in the prompt emission jet, likely the maximum \citep[e.g.][]{Beniamini15}, which demands the least energy from the central engine. The biggest unknown when placing limits on the jet opening angle is the density of the surrounding medium. Our most constraining limit comes from the compactness of the source that is required to produce an amplitude of scintillation sufficient to match the radio observations, since a denser medium inhibits the growth of the fireball. In our scintillation calculations, we have minimised compactness, and therefore density, by accepting the case in which each of the data points are at the maximum amplitude of scintillation away from their true value. If one, or both, data points are in fact not maximally scintillated, then the source must be even more compact, requiring the density to be higher, and more energy to prevent a jet break in the X-ray light curve. This is already assuming that the synchrotron self-absorption frequency lies between $9$ and $18$~GHz.

The only assumption that is not stacked in favour of minimising the energy demands from the central engine is the scintillation measure. The \citet{Cordes02} free election distribution model gives log SM $= -3.75$~m$^{-20/3}$~kpc, but has a scatter at high galactic latitudes of roughly a factor of three. An upper limit of log SM $= -3.45$~m$^{-20/3}$~kpc could therefore be used when calculating the amplitude of scintillation to reduce the compactness demands further, lowering the density limits. If this concession is made, the jet opening angle can be reduced to $\theta_0 \geq 0.09$~radians in the wind case, or $\theta_0 \geq 0.21$~radians for the ISM. This drops the energy demands down to $E_{\rm wind} \geq 4.62 \times 10^{51}$~erg, or $E_{\rm ISM} \geq 2.51 \times 10^{52}$~erg; if every assumption is fine-tuned in favor of saving energy, the absolute minimum energy demand on the system can be dropped to $6.6 \times 10^{51}$~erg (minimum GRB + minimum SN); just barely inside the capabilities of a $2$~ms magnetar. However, this requires:
\begin{itemize}
\item Maximum ($50$ per cent) efficiency.
\item The maximum amplitude of scintillation in each radio data point.
\item The modelled scintillation measure in the direction of GRB 111209A to be maximally inaccurate (in the favourable direction).
\item $9$~GHz $< \nu_a < 18$~GHz.
\item $\epsilon_B$ to be at the high end of the observed distribution.
\item The true density to sit right at the lower limit, so that a jet break occurred in GRB 111209A almost immediately after \emph{Swift} stopped observing it.
\item The medium surrounding GRB 111209A to be a high density wind.
\item SN 2011kl to be aspherical to a degree comparable to SN 1998bw.
\end{itemize}
Though some of these conditions may be true, it seems extremely unlikely that all of them are. We therefore consider it highly likely that if a magnetar is responsible for GRB 111209A, it must have had a spin period not much slower than $1$~ms, and likely would have had to have occurred in a wind-like medium.

If an underlying magnetar in GRB 111209A must have a spin period close to $1$~ms, we must also assess how this fits with the observed supernova peak time and luminosity, since the central engine must account for both. SN 2011kl had a peak time of $23.5$ observer frame days and a peak luminosity of $2.8^{+1.2}_{-1.0}\times10^{43}$~erg~s$^{-1}$ \citep{Greiner15}. Using the simplified analytical model of a super-luminous supernova \citep{Kasen10}, we find the data can be best matched by an ejecta mass in the range of $1.4$~M$_{\sun}$ to $2.8$~M$_{\sun}$ for a 1~ms magnetar, depending on how much energy is in the SN. GRB-SNe typically have ejecta masses of around $10$~M$_{\sun}$ \citep[e.g.][]{Mazzali03,Deng05}, but other studies into SN 2011kl have found equally low ejecta masses \citep{Greiner15,Metzger15,Bersten16,Cano16}, and $2$~M$_{\sun}$ has been inferred previously in other GRB-SNe \citep[e.g. SN 2006aj;][]{Mazzali06}. GRB 060218, associated with SN 2006aj, also had an early period of emission that lasted several thousands of seconds (when XRT data are included), and has a phenomenology not unlike GRB 111209A. This burst was not seen by \emph{Konus-Wind}, despite the hardness ratio observed by \emph{Swift}-BAT being very similar to that seen in GRB 111209A. It may be that the $T_{90}$ obtained from low Earth orbit is not robust, and is only a fraction of the true duration of GRB 060218; in GRB 111209A, the BAT-measured $T_{90}$ is only $320$~s, compared to the $10^4$~s measured by \emph{Konus-Wind}. If GRB 060218 was in fact ultra-long, it could be that low ejecta masses are characteristic of the population, perhaps due to the division of the energy between the GRB and the SN. However, it's worth noting that at a redshift of only $0.033$ \citep{Mirabal06}, $E_{\rm \gamma ,iso} \approx (1.9 \pm 0.1) \times 10^{49}$~erg \citep{Sakamoto06b}, significantly less than for GRB 111209A.

ULGRB 111209A shows evolution in its X-ray hardness ratio, transitioning from around $1$ at $10^3$~s to close to $3$ at $10^6$~s. This hard-to-soft evolution is consistent with the dust scattering model \citep[e.g.][]{Shen09}, in which hard X-ray photons are reprocessed by a screen of dust several tens of parsecs from the GRB site. Reprocessed emission may be capable of masking a jet break, since the emission will be dominated by the reflected photons. The size of the hardness ratio shift is comparable to that seen in another ULGRB, 130925A, which went from a hardness ratio of about $2$ to close to $4$ over a similar time scale. \citet{Evans14} showed that 130925A was a good candidate for dust scattering, but when assessing 111209A, they find that the early light curve is dominated by fading prompt emission until around $10^5$~s after trigger. When fitting the afterglow from $10^5$~s onwards, \citet{Evans14} find the evolution is only significant to the $1.5\sigma$ level, and therefore conclude that there is no concrete evidence for scattering in this case. Since the dust echo interpretation is not statistically significant, we conclude that the emission observed comes directly from the GRB afterglow, and the lack of an observed jet break is still constraining to the jet opening angle, and therefore the energy of the burst.

Although the wind environment is a better fit with the energetic limits of a magnetar central engine, the high strength wind implied for GRB 111209A is problematic; GRBs are believed to require a rapidly-rotating progenitor star \citep{Levan16}, but a strong wind carries away angular momentum, slowing the spin. In addition, wind power correlates with metallicity, but GRBs favour low-metallicity environments \citep[e.g.][]{Levesque10d}, and GRB 111209A itself was found to occur in a particularly low-metallicity galaxy, with $Z = 7.95^{+0.30}_{-0.17}$~12+log(O/H) \citep{Kruhler15}.

One of the fundamental properties that drives the evolution of GRB afterglows is the power law index of the electron energy distribution, $p$. This single value can be related to the spectral and temporal slopes of the afterglow at different segments of the synchrotron spectrum (see Figure~\ref{fig:SED}) using the synchrotron closure relations \citep[e.g.][]{Gao13}. The way that the spectral and temporal indices relate to the underlying electron energy distribution depends on the type of environment the fireball is expanding into, and so we can use our measurements at X-ray and optical frequencies as an indicator of whether an ISM or wind-like environment is preferred by seeing which gives a better agreement for a single value of $p$. We find that an ISM environment gives better agreement between the measured indices, with a best fit $p = 2.73 \pm 0.05$ ($\chi^2$/dof = 4.9 for a $\chi^2$ of $14.7$ and 3 degrees of freedom), versus $p = 2.42 \pm 0.05$ ($\chi^2$/dof = 19.0 for a $\chi^2$ of 57.0 and 3 degrees of freedom). However, neither of these can be considered a good fit, and lacking concrete observational evidence on the environment type, the only conclusion is that the closure relations `disagree less' with the ISM environment than the wind. This is not infrequently the case in GRBs \citep[e.g.][]{Wang15}, as the standard synchrotron model is only a relatively simple analytical approximation of a very complex process.

If GRB 111209A is in an ISM environment, the maximum beaming factor required to avoid a jet break is around $21.0$. This wide jet is in agreement with the opening angle inferred by \citet{Ryan15}, who obtained their values by fitting afterglow light curves to hydrodynamic simulations using the {\sc ScaleFit} package, though it should be noted that the wind-like environment was not considered in their study. If the jet is indeed this wide, the beaming-corrected energy cannot be below about $5 \times 10^{52}$~erg, perilously close to the maximum available energy from a ms NS.

If the progenitor of GRB 111209A/SN 2011kl is not a magnetar, but is in fact a black hole, then the energy limit in Equation~\ref{eqn:energy} does not apply, and the jet opening angle can be significantly wider. The lack of a jet break would then not be a problem. However, since the peak luminosity of SN 2011kl requires an emission mechanism in addition to $^{56}$Ni, some additional component must be invoked in order to power it. \citet{Gao16} model GRB 111209A/SN 2011kl as a collapsar. To explain the peak of SN 2011kl, they use fallback accretion, which drives a late outflow that delivers energy to the SN. Additionally, models invoking the tidal disruption of a white dwarf by a black hole have been discussed \citep{Ioka16}.

\section{Conclusions}\label{sec:conclusions}

We have shown that the combined energy requirements of ULGRB 111209A and the associated SN 2011kl demand a magnetar with a spin period not much greater than $1$~ms unless the CBM density is very low. However, we have provided two independent indicators, the frequency of the synchrotron cooling break in the GRB spectrum, and the likely strong scintillation of the radio observations of the burst, which show that this is not the case. We therefore conclude that ULGRB 111209A/SN 2011kl exploded in a dense environment \citep[in agreement with][]{Gao16}, and can only have been powered by a neutron star with a spin period of $\sim 1$~ms or faster to be energetically consistent with the magnetar model. The energy demand if the local environment is wind-like in its density profile sits within magnetar limits at a few $10^{52}$~erg, but an ISM-like medium pushes the model to more extreme (though still theoretically attainable) energies above $5 \times 10^{52}$~erg. Observations of this event do not provide strong evidence either way for the density profile of the surrounding medium.

One of our key conclusions is that extended follow up of ULGRBs at X-ray frequencies could provide a stern test of the magnetar central engine, because even at around three weeks of coverage, the energy output of GRB 111209A came close to the magnetar limit. In the wind environment, $E_{\rm \gamma ,iso} \propto t_j$, whereas in the ISM environment, $E_{\rm \gamma ,iso} \propto t_j^3$, which means in either case, a factor of two in the duration of coverage would have been significant. Follow-up observations with \emph{Chandra} or XMM-\emph{Newton} could therefore expect to find a jet break in a similar GRB if a magnetar is present, or drive the energy output required to values high enough to rule out the magnetar central engine. We also argue that enhanced radio coverage is vital, as this would have allowed us to tie down the influence of scintillation more precisely. With just a modest improvement in the dataset for a future GRB, we have a metric capable of distinguishing between the competing central engine models.

\section{Acknowledgements}

We thank Alexander van der Horst for useful comments and feedback that improved the manuscript. We also thank Paul Hancock, Phil Evans and Paul O'Brien for helpful discussions. We are grateful to the anonymous referee, whose comments and suggestions enhanced the manuscript. This work made use of data supplied by the UK \emph{Swift} Science Data Centre at the University of Leicester.

\bibliographystyle{aasjournal}
\bibliography{../../bpg6/papers/ref}

\end{document}